\documentclass{aa}  
\usepackage{graphicx}
\usepackage{txfonts}
\begin{document}
%


\title{A mean redshift of 2.8 for {\it Swift} gamma-ray bursts}

\author{
P.~Jakobsson \inst{1} \and
A.~Levan  \inst{2,3} \and
J.~P.~U.~Fynbo \inst{1} \and
R.~Priddey \inst{2} \and
J.~Hjorth \inst{1} \and
N.~Tanvir \inst{2} \and
D.~Watson \inst{1} \and
B.~L.~Jensen \inst{1} \and
J.~Sollerman \inst{1}\and
P.~Natarajan \inst{4} \and
J.~Gorosabel \inst{5} \and
J.~M.~Castro~Cer\'on \inst{1} \and
K.~Pedersen \inst{1}\and
T.~Pursimo \inst{6} \and
A.~S.~\'Arnad\'ottir \inst{3} \and
A.~J.~Castro-Tirado \inst{5} \and
C.~J.~Davis \inst{7} \and
H.~J.~Deeg \inst{8} \and
D.~A.~Fiuza \inst{9} \and
S.~Mykolaitis \inst{10} \and
S.~G.~Sousa \inst{11,12}
}

\institute{
          Dark Cosmology Centre, Niels Bohr Institute, University of
          Copenhagen, \mbox{Juliane Maries Vej 30, 2100 Copenhagen, Denmark}
          \and
          Centre for Astrophysics Research, University of Hertfordshire,
          College Lane, Hatfield, Herts, AL10 9AB, UK
   	  \and 
	  Lund Observatory, Box 43, SE-221 00, Lund, Sweden 
          \and
          Department of Astronomy, Yale University, PO Box 208101, 
          New Haven CT 06520-8101, USA
	  \and
          Instituto de Astrof\'{\i}sica de Andaluc\'{\i}a (CSIC), 
          Apartado de Correos, 3004, E-18080 Granada, Spain
          \and
          Nordic Optical Telescope, Apartado de Correos, 474, E-38700 
          Santa Cruz de la Palma (Tenerife), Spain
          \and
          Joint Astronomy Centre, University Park, 660 North A'ohoku Place, 
          Hilo, HI 96720, USA
          \and
          Instituto de Astrof\'{\i}sica de Canarias, c/. V\'{\i}a L\'actea, 
          s/n, E-38200 La Laguna (Tenerife), Spain
          \and
	  Observatory, University of Helsinki, PO Box 14,
          FIN-00014 University of Helsinki, Finland
          \and
          Institute of Theoretical Physics and Astronomy,
          A. Gostauto St. 12, 01108 Vilnius, Lithuania
	  \and
          Centro de Astrof\'{\i}sica da Universidade do Porto, 
          Rua das Estrelas, P-4150-762 Porto, Portugal
          \and
          Centro de Astrof\'{\i}sica Observat\'orio Astron\'omico de Lisboa, 
          Tapada da Ajuda, P-1349-018 Lisboa, Portugal
}


\date{Received 29 September 2005 / Accepted 21 October 2005}
\abstract{
The exceptionally high luminosities of gamma-ray bursts (GRBs), gradually 
emerging as extremely useful probes of star formation, make them promising 
tools for exploration of the high-redshift Universe. Here we present a 
carefully selected sample of \emph{Swift} GRBs, intended to estimate in an
unbiased way the GRB mean redshift ($z_\mathrm{mean}$), constraints on the 
fraction of high-redshift bursts and an upper limit on the fraction of 
heavily obscured afterglows. We find that $z_\mathrm{mean} = 2.8$ and that 
at least 7\% of GRBs originate at $z > 5$. In addition, consistent with 
pre-\emph{Swift} observations, at most 20\% of afterglows can be heavily 
obscured. The redshift distribution of the sample is qualitatively consistent 
with models where the GRB rate is proportional to the star formation rate in 
the Universe. We also report optical, near-infrared and X-ray observations of 
the afterglow of GRB\,050814, which was seen to exhibit very red optical 
colours. By modelling its spectral energy distribution we find that 
$z = 5.3 \pm 0.3$. The high mean redshift of GRBs and their wide redshift 
range clearly demonstrates their suitability as efficient probes of galaxies 
and the intergalactic medium over a significant fraction of the history of 
the Universe.


}



\maketitle

\section{Introduction}
The potential of long-duration ($>$2\,s) gamma-ray bursts (GRBs) as probes of 
the high-redshift Universe has long been recognised. The immense luminosities 
of the bursts, coupled with their origin in the core collapse of massive stars 
(Hjorth et al. \cite{jensNature}; Stanek et al. \cite{stanek}) and 
$\gamma$-ray penetration through dust, suggests a variety of intriguing 
applications. Much effort has been directed into the use of bursts for 
studying star formation (e.g. Fruchter et al. \cite{andy}; Christensen et al. 
\cite{lise}; Tanvir et al. \cite{tanvir04}; Jakobsson et al. \cite{palli05}), 
as backlights for exploring high-redshift galaxies and the intergalactic 
medium (e.g. Vreeswijk et al. \cite{paul}; Jakobsson et al. \cite{palli04}) 
and even as probes of cosmological parameters (e.g. Ghirlanda et al. 
\cite{ghirlanda}; M\"ortsell \& Sollerman \cite{mortsell}). Although the GRB 
population observed until the end of 2004 had enabled much progress in the 
field, it was widely expected that the launch of {\it Swift}, and the 
subsequent order of magnitude increase in the number of GRBs open to detailed 
study, would allow further insight into the high-redshift Universe (Gehrels 
et al. \cite{gehrels}). Indeed, the ability of {\it Swift} to locate and 
follow-up a fainter burst population than was previously possible (Berger et 
al. \cite{edo05}) has allowed more distant bursts to be studied. The mean 
redshift of pre-{\it Swift} bursts was $z_\mathrm{mean} = 1.4$, while bursts 
discovered by {\it Swift} now have $z_\mathrm{mean} = 2.8$, including the 
first burst to have been discovered with $z > 6$, GRB\,050904 at $z=6.29$ 
(Haislip et al. \cite{haislip}; Kawai et al. \cite{kawai}; Price et al. 
\cite{price}; Tagliaferri et al. \cite{tag}: Watson et al. \cite{darach904}).
\par
High-redshift bursts, whose afterglows may already be faint due to the 
large luminosity distance\footnote{However, due to time dilation, little 
decrease is expected in the spectral energy flux in a given frequency band 
and at a fixed time of observation after the GRB with increasing redshift
(Lamb \& Reichart \cite{lamb}).} (e.g. GRB\,020124: Berger et al. 
\cite{edo02}; Hjorth et al. \cite{hjorth}) are rendered essentially 
invisible at optical wavelengths for $z > 6$ due to hydrogen opacity 
when the redshifted Ly$\alpha$ break sweeps through the optical regime; an 
effect that is clearly seen in galaxies in the Hubble Deep Field 
(e.g. Spinrad et al. \cite{spinrad}; Weymann et al. \cite{weymann}).
Therefore, a simple diagnostic of a high-redshift burst is its absence in 
deep optical observations and detection in a redder filter. Such a 
detection does not unambiguously fix the redshift, since high local 
extinction ($A_V$) will also markedly reduce the short wavelength flux.
However, with multi-wavelength observations it is possible to 
distinguish the curved red spectrum expected for an extinguished burst, 
from the sharp cut-off due to the Ly$\alpha$ break. This method is similar 
to that which has been used successfully (and accurately) in the selection 
and study of Lyman-break galaxies (Steidel \& Hamilton 
\cite{steidel92,steidel93}), but in fact is likely to be even more robust 
thanks to the simple power-law spectra exhibited by GRB afterglows.
\par
In this paper we introduce an objective {\it Swift} sample, appropriate 
for determining the $z_\mathrm{mean}$ of GRBs, estimate the fraction of 
high-redshift bursts and an upper limit on the fraction of heavily obscured 
afterglows. We also compare the redshift distribution of the sample 
to models, predicted before the availability of these results, where the 
GRB rate is proportional to the star formation rate. In addition, we present 
optical, near-infrared (NIR) and X-ray observations of the GRB\,050814 
afterglow and a fit to the resulting spectral energy distribution (SED), 
allowing us to determine a well constrained redshift of $z = 5.3 \pm 0.3$. 
\section{Observations}
\label{obs.sec}
GRB\,050814 was discovered by the Burst Alert Telescope (BAT) aboard the 
{\it Swift} satellite on 14.485 August 2005 UT (Retter et al. \cite{retter}). 
The burst exhibited a slow rise and decline above the background level, with 
a poorly constrained $t_{90}$ of $65^{+40}_{-20}$\,s (Tueller et al. 
\cite{tueller}). The X-ray Telescope (XRT) slewed prompt\-ly to the location 
and began taking data at \mbox{$\Delta t = 138$\,s}, where $\Delta t$ 
is the time from the onset of the burst, revealing a fading X-ray source 
(Morris et al. \cite{morris}). Observations with the UV/Optical Telescope
began at \mbox{$\Delta t = 167$\,s} but failed to reveal an optical afterglow
(OA) candidate to a limiting magnitude of $V = 20.5$\,mag (Blustin et al. 
\cite{blustin}). A marginally (2$\sigma$) detected OA candidate was reported
by Cenko (\cite{brad}) based on observations at the Palomar 60-inch 
telescope.
\par
We observed the GRB\,050814 XRT error circle in $VRI$ with the 
Andaluc\'{\i}a Faint Object Spectrograph and Camera (ALFOSC) on the Nordic 
Optical Telescope (NOT) at \mbox{$\Delta t \approx 13.5$\,hr}. A point-like 
object was detected displaying a red $R-I$ colour, its absence in the 
$V$-band suggesting a high-redshift origin (Jensen et al. \cite{brian}). 
Additional $I$-band imaging was secured at the NOT using both ALFOSC and the 
MOSaic CAmera (MOSCA), confirming the fading behaviour of the 
OA candidate. Supplementary $R$-band imaging was obtained at the 
Isaac Newton Telescope (INT) using the Wide Field Camera (WFC). All the 
data were de-biased, flatfield corrected and combined using standard methods. 
The instrumental optical magnitudes were transformed to the Johnson 
photometric system using observations of Landolt (\cite{landolt}) standards. 
The journal of the observations including the results of our photometry is 
presented in Table~\ref{phot.tab}.
\begin{table}[]
\caption[]{A log of the GRB\,050814 follow-up imaging observations.
Limiting magnitudes are $2 \sigma$  in a circular aperture with a
radius equal to the seeing. No correction for Galactic extinction has 
been applied to the photometry. Note that the NOT and INT $R$-band
filters are different, resulting in a slight magnitude discrepancy
due to the Ly$\alpha$ break presence in this band.}
\label{phot.tab}
\begin{minipage}{\columnwidth}
\centering
\setlength{\arrayrulewidth}{0.8pt}   
\begin{tabular}{rcccr@{}}
\hline
\hline
\vspace{-2 mm} \\
$\Delta t$\footnote{Days after 14.485 August 2005 UT.} \hspace{0.8 mm} 
& Telescope/ & Magnitude & Seeing  & Exp. time \\
\hspace{-0 mm}[days] & Instrument &   & [arcsec] & [s] \hspace{3.8 mm} \\
\vspace{-2 mm} \\
\hline
\vspace{-2 mm} \\
\hspace{-3 mm}
\emph{V-band:}&&&&\\
0.555 & NOT/ALFOSC  &     $>25.2$      & 0.9 & $3 \times 300$       \\
\hspace{-3 mm}
\emph{R-band:}&&&&\\    
0.432 & INT/WFC     & $23.09 \pm 0.21$ & 1.2 & $300$                \\
0.488 & INT/WFC     & $23.09 \pm 0.23$ & 1.4 & $300$                \\
0.569 & NOT/ALFOSC  & $23.42 \pm 0.09$ & 0.9 & $3 \times 300$       \\
0.584 & INT/WFC     & $23.09 \pm 0.23$ & 1.6 & $300$                \\
\hspace{-3 mm}
\emph{I-band:}&&&&\\    
0.584 & NOT/ALFOSC  & $20.55 \pm 0.04$ & 0.9 & $600$                \\
1.475 & NOT/ALFOSC  & $21.90 \pm 0.08$ & 0.8 & $2 \times 500$       \\
20.375 & NOT/MOSCA  &     $>24.7$      & 0.8 & $9 \times 400$       \\
\hspace{-3 mm}
\emph{J-band:}&&&&\\    
2.769 & UKIRT/UTFI  & $22.40 \pm 0.38$  & 0.7 & $64 \times 30$       \\
\hspace{-3 mm}
\emph{K-band:}&&&&\\    
0.935 & UKIRT/UTFI  & $17.60 \pm 0.03$ & 0.6 & $45 \times 20$       \\
2.737 & UKIRT/UTFI  & $20.02 \pm 0.12$ & 0.7 & $80 \times 20$       \\
\vspace{-2 mm} \\
\hline
\end{tabular}
\end{minipage}
\end{table}     
\par
NIR observations were pursued at the United Kingdom Infrared 
Telescope (UKIRT) using the UKIRT Fast Track Imager (UFTI) at 
\mbox{$\Delta t \approx 22$\,hr} ($K$) and 
\mbox{$\Delta t \approx 66$\,hr} ($JK$). The data were pipeline processed 
through the {\sc ORACDR} system (Cavanagh et al. \cite{cavanagh}) 
to produce flattened, dark- and sky-subtracted mosaic images. Flux
calibration was performed against the standard star FS141.
\par
We have also analysed the \emph{Swift}/XRT observations of GRB\,050814 and 
used these to complement our optical/NIR dataset. Processed event files from
the archive were used and background subtracted spectra and light curves
from Windowed Timing (WT) and Photon Counting (PC) modes were produced in
a standard way.
\section{Spectral energy distribution of the afterglow}
\label{sed.sec}
Our multiband observations of the GRB\,050814 afterglow allowed the 
construction of its SED. We extrapolated the magnitudes to a common epoch, 
namely the nearly simultaneous epoch of our initial $VRI$ NOT observations 
($\Delta t = 14$\,hr). This was carried out by taking into account the decay 
indices (as calculated from our photometry in Table~\ref{phot.tab})
\mbox{$\alpha_I = 1.34 \pm 0.12$} and \mbox{$\alpha_K = 2.08 \pm 0.12$}
($F_{\nu} \propto t^{-\alpha_{\nu}} \nu^{-\beta}$). This discrepancy in the 
decay indices implies a late time ($\Delta t \gtrsim 24$\,hr) rapid decay 
with \mbox{$\alpha > 2.08$}, although the sparse sampling of the light curve 
\mbox{means that} the timing of this break is poorly constrained. It might
be the jet break, a suggestion supported by the fact that the XRT light 
curve displays a break at approximately 28\,hr with a pre-break slope
of \mbox{$\alpha_\mathrm{X} = 0.58 \pm 0.05$}. We have used 
$\alpha_\mathrm{X}$ to obtain lower limits on the extrapolated $JK$ fluxes.
\par
\begin{figure}
\centering
\resizebox{\hsize}{!}{\includegraphics[bb=92 371 540 696,clip]{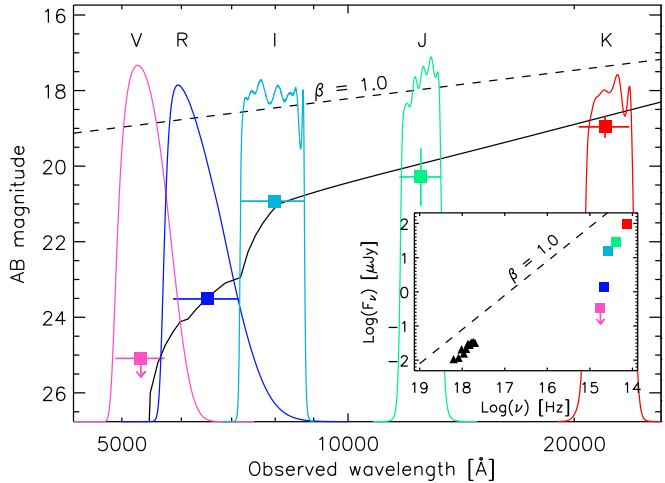}}
\caption{The spectral energy distribution of the GRB 050814 afterglow at
$\Delta t = 14$\,hr. The strong break blueward of the $I$-band is 
too strong to be readily explained by reddening alone and is best fit by 
the presence of the Ly$\alpha$ break at $z = 5.3$. The solid curve is a
fit to the data at that redshift. The dashed line shows the spectral 
slope expected from a synchrotron emission in the fireball
model with $\beta = 1$. The filter response functions are also shown.
The horizontal error bars represent the FWHM of each filter. The
$V$-band upper limit is 2$\sigma$. The inset shows the $VRIJK$ observations
(filled squares) along with the X-ray spectrum (filled triangles) at
$\Delta t = 14$\,hr. The dashed line is the same $\beta = 1$ slope as
in the main panel.}
\label{sed.fig}
\end{figure}

The SED is shown in Fig.~\ref{sed.fig}, where we have corrected the 
observed data points for foreground (Galactic) extinction using the 
reddening maps of Schlegel et al. (\cite{schlegel}), giving 
\mbox{$E(B-V) = 0.028$}\,mag at that position on the sky. \mbox{We note} 
that the flux from the host galaxy has not been subtracted. However, the host 
is faint enough ($I > 24.7$\,mag) that it should contribute less than $2$\% 
of the flux at the epoch we are exploring. The SED has a strong break blueward 
of the $I$-band, exhibiting colours of \mbox{$I - K = 3.44 \pm 0.29$\,mag} and 
\mbox{$R - I = 2.87 \pm 0.10$\,mag}, corresponding to spectral slopes of 
\mbox{$\beta_{IK} = 1.78 \pm 0.12$} and \mbox{$\beta_{RI} = 11.70 \pm 0.04$}, 
respectively. The latter value is unreasonable for GRB afterglows, implying 
an electron energy power-law index more than ten times higher than 
normally observed. Even in the case of high $A_V$, such steep 
slopes cannot be obtained (see below and also Reichart \cite{reichart}). 
\par
\begin{figure}
\centering
\resizebox{\hsize}{!}{\includegraphics[bb=72 371 535 696,clip]{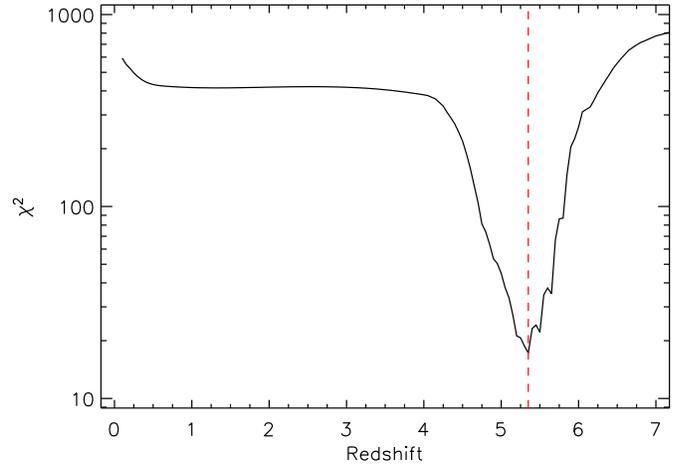}}
\caption{The calculated $\chi^2$ as a function of the GRB\,050814 redshift. 
Low redshifts are ruled out by the data and cannot be fit by extinction. 
Very high $z > 7$ redshifts are ruled out by the presence of the 
afterglow in the $I$-band. The best fit redshift, indicated by the
vertical dashed line, is $z=5.3$ although a range of redshifts from 
$5.0 < z < 5.6$ have similar $\chi^2$ values.}
\label{chi.fig}
\end{figure}

The most likely explanation for the steep break observed is due to the 
presence of the Ly$\alpha$ break at a redshift of $5 < z < 6$. To provide a 
more robust estimate of the GRB\,050814 redshift we fit the available 
photometry at different redshifts, allowing for a range in $\beta$ and
$A_V$ modelled using the parametrization of Calzetti et al. 
(\cite{calzetti}). The models of Madau (\cite{madau}) provide the average 
hydrogen opacity as a function of redshift. Figure~\ref{chi.fig} shows the 
minimum $\chi^2$ for each redshift step plotted against redshift; the best
fit is obtained for $z = 5.3 \pm 0.3$ and is shown in Fig.~\ref{sed.fig}. 
However, we are only able to obtain weak constraints on $\beta$ and $A_V$. 
Fixing $\beta = 1.0$, a typical value for GRB afterglows, results in a best 
fit of a restframe $A_V = 0.9$\,mag and an unchanged redshift. This $A_V$ is 
margin\-ally higher than has been inferred from the SEDs of pre-Swift bursts 
with bright OAs 
($A_V = 0.09$\,mag in GRB\,000301C: Jensen et al. \cite{brian01};
$A_V = 0.18$\,mag in GRB\,000926: Fynbo et al. \cite{johan01};
$A_V = 0.08$\,mag in GRB\,011211: Jakobsson et al. \cite{palli03};
$A_V < 0.20$\,mag in GRB\,020124: Hjorth et al. \cite{hjorth};
$A_V = 0.26$\,mag in GRB\,021004: Holland et al. \cite{stephen};
$A_V < 0.50$\,mag in GRB\,030323: Vreeswijk et al. \cite{paul};
$A_V = 0.34$\,mag in GRB\,030429: Jakobsson et al. \cite{palli04}) but is a 
necessary consequence of the red $I - K$ colour. The extrapolated 
$\beta = 1.0$ line, normalized for $A_V = 0.9$\,mag, slightly overestimates 
the predicted X-ray flux (inset of Fig.~\ref{sed.fig}), indicating that 
$\beta$ is a bit steeper; $\beta = 1.1$ would make the X-ray data fall on 
the line. Since the best fit X-ray spectral index is consistent with the
assumed \mbox{optical/NIR} one, a cooling break between the optical and 
X-rays can be ruled out.
\par
\begin{table*}[]
\caption[]{A list of all long-duration GRBs which have a Galactic extinction 
$A_V^\mathrm{Gal} < 0.5$\,mag, a declination between $-70^{\circ}$ and
$+70^{\circ}$, and are localised with the \emph{Swift}/XRT after 
\mbox{1 March} 2005. Here $\theta_\mathrm{Sun}$ is the Sun-to-field 
distance, $\theta_\mathrm{Moon}$ the Moon-to-field distance and 
I$_\mathrm{Moon}$ the Moon illumination at the time the burst occurred. For 
a burst detected in the optical but without a reported redshift, an upper 
redshift limit is estimated based on the filter it is detected in. 
References are 
[1] Kelson \& Berger (\cite{kelson});
[2] Berger \& Mulchaey (\cite{berger318});
[3] Fynbo et al. (\cite{johan319});
[4] Watson et al. (\cite{darach});
[5] Rol et al. (\cite{rol});
[6] Cenko et al. (\cite{cenko416A});
[7] Cenko et al. (\cite{cenko502B});
[8] Berger et al. (\cite{berger505});
[9] Foley et al. (\cite{foley});
[10] Berger \& Becker (\cite{berger603});
[11] Tanvir et al. (\cite{tanvir});
[12] Poole et al. (\cite{poole});
[13] Starling et al. (\cite{rhaana});
[14] Blustin et al. (\cite{blustin801});
[15] Fynbo et al. (\cite{johan802});
[16] This work;
[17] Prochaska et al. (\cite{jason});
[18] Fynbo et al. (\cite{johan824});
[19] Kawai et al. (\cite{kawai});
[20] Fugazza et al. (\cite{fugazza});
[21] Bloom (\cite{bloom});
[22] Jakobsson et al. (\cite{palli-z}).
}
\label{sample.tab}
\begin{minipage}{18.0cm}
\centering
\setlength{\arrayrulewidth}{0.8pt}   
\begin{tabular}{@{}lccrrrr|lccrrrr@{}}
\hline
\hline
GRB & $z$ & $A_V^\mathrm{Gal}$ & $\theta_\mathrm{Sun}$ & 
$\theta_\mathrm{Moon}$ & I$_\mathrm{Moon}$ & Ref. &
GRB & $z$ & $A_V^\mathrm{Gal}$ & $\theta_\mathrm{Sun}$ & 
$\theta_\mathrm{Moon}$ & I$_\mathrm{Moon}$ & Ref. \\
    &     & [mag] & [deg]                  & [deg]               & [\%] & &
    &     & [mag] & [deg]                  & [deg]               & [\%] & \\
\hline
050315  & 1.95 & 0.16 & 59  & 114 & 24  & 1 &
050730  & 3.97 & 0.17 & 84  & 150 & 31  & 13 \\

050318  & 1.44 & 0.06 & 64  & 83  & 52  & 2 &
050801  & $<2.0$ \hspace{3.7mm} & 0.32 & 82  & 127 & 15 & 14 \\

050319  & 3.24 & 0.04 & 132 & 44  & 61  & 3 &
050802  & 1.71 & 0.07 & 79  & 102 & 9   & 15 \\

050401  & 2.90 & 0.22 & 122 & 36  & 62  & 4 &
050803  &      & 0.25 & 136 & 113 & 4   &    \\

050406  & $<3.5$ \hspace{3.7mm} & 0.07 & 59  & 58  & 10  & 5 &
050814  & 5.3  & 0.09 & 99  & 74  & 60  & 16  \\

050412  &      & 0.07 & 159 & 121 & 11  &   &
050819  &      & 0.40 & 132 & 59  & 99  &     \\

050416A & 0.65 & 0.10 & 145 & 69  & 44  & 6 &
050820A & 2.61 & 0.15 & 147 & 34  & 100 & 17 \\

050502B\footnote{The OA was detected in $I$ but not $V$, suggesting a
high redshift (Cenko et al. \cite{cenko502B}).} 
& $<8.5$ \hspace{3.7mm} & 0.10 & 98  & 177 & 42  & 7 &
050822  &      & 0.05 & 105 & 63  & 93  &    \\

050505  & 4.27 & 0.07 & 90  & 130 & 13  & 8 &
050824  & 0.83 & 0.12 & 129 & 17  & 77  & 18 \\

050525  & 0.61 & 0.32 & 121 & 57  & 98  & 9 &
050904  & 6.29 & 0.21 & 143 & 145 & 0   & 19 \\

050603  & 2.82 & 0.09 & 56  & 39  & 15  & 10 &
050908  & 3.34 & 0.08 & 146 & 151 & 16  & 20 \\

050714B &      & 0.18 & 67  & 25  & 44  &    &
050915A & $<13.0$ \hspace{5.2mm}     & 0.09 & 93  & 109 & 86  & 21 \\

050716  & $<8.0$ \hspace{3.7mm} & 0.37 & 108 & 125 & 64  & 11 &
050922B &      & 0.12 & 171 & 57  & 80  &    \\

050726  & $<5.0$ \hspace{3.7mm} & 0.21 & 88  & 147 & 73 & 12 &
050922C & 2.20 & 0.34 & 138 & 93  & 80  & 22 \\

\hline
\end{tabular}
\end{minipage}
\end{table*}     
We have also explored whether the observed SED could be explained by a lower 
redshift and a large extinction assuming the three extinction laws given by
Pei (\cite{pei}) for the Milky Way (MW), Large Magellanic Cloud (LMC)
and Small Magellanic Cloud (SMC). This was done in the following way.
We have calculated the observed $E(R-I) \equiv A_R - A_I = 2.41 \pm 0.10$\,mag
based on $\beta = 1.0$ (see above). By (blue)shifting the $R$ and $I$ filters
along the aforementioned extinction laws, we can estimate which 
\mbox{redshifts} produce the observed $E(R-I)$. Using these redshifts we can 
then compute the predicted $E(I-K)$ and compare to the observed value of 
$0.88 \pm 0.29$\,mag. In every case the predicted values are 
$E(I-K) > 3.0$\,mag, implying that none of the extinction laws can give rise 
to the observed SED at a low redshift. In other words, the spectral break 
between the observed $R$- and $I$-bands combined with the shallower slope 
between $I$ and $K$ is inconsistent with the MW, SMC and LMC extinction 
laws at any lower redshift. Dust close to the GRB site might not follow 
these extinction laws, for example due to dust being destroyed by the prompt 
radiation of the GRB itself (e.g. Perna et al. \cite{perna}). However, that 
would tend to produce greyer dust (small grains being preferentially removed), 
thus reducing rather than increasing the amount of reddening.
\par 
The X-ray limits on $N_{\rm H}$ from the XRT also support a high-redshift
origin for the red optical/NIR SED, by providing an upper limit on the 
possible absorbing extragalactic oxygen column density along 
the line of sight. The values of $N_{\rm H}$ we derive from the fitting 
process are sensitive to the fit at low energies. We have used ancillary 
response files for the WT mode data, generated by \texttt{xrtmkarf} with 
``inarffile=NONE'', which are better at reproducing the low energy 
continua. The $N_{\rm H}$ limits using this response file are slightly 
higher than using the default ancillary response.
\par
We can then obtain a limit on $A_V$ using a Galactic empirical relation stating
that $A_V = 1$\,mag corresponds to $N_{\rm H} = 1.79\times10^{21}$\,cm$^{-2}$ 
(Predehl \& Schmitt \cite{predehl}). This implies that for redshifts of 
$z = (0.5, 1.0, 2.0, 3.0)$ the respective 3$\sigma$ upper limits on $A_V$ 
are (0.5, 0.8, 1.8, 3.6)\,mag. None of these are \mbox{large enough} to 
simultaneously reproduce the observed $R - I$ and $I - K$ colours at
low redshifts.
\section{The redshift distribution of {\it Swift} bursts}
\label{sample.sec}
At the time of this writing (30 September 2005) {\it Swift} has been 
operating for approximately 9 months and has detected a total of 70
long-duration GRBs. In order to study the redshift distribution of GRBs, it 
is important to carefully select a sample containing bursts which have 
``observing conditions'' favorable for redshift determination. Our first 
criteria is small error circles, hence the bursts have to be localised with 
the XRT (56 GRBs). In addition we require the Galactic extinction in the 
direction to the burst to be sufficiently small or $A_V^\mathrm{Gal} < 
0.5$\,mag (38 GRBs). Thirdly, the XRT error circle should be distributed 
quickly for a relatively rapid follow-up. Although the automatic slewing of 
{\it Swift} was enabled in the middle of January 2005, part of the following 
month was dedicated to calibration which could not be interrupted. Therefore, 
we have only included bursts occurring after 1 March 2005 (31 GRBs). Finally,
we have rejected (3) bursts with a declination unsuitable (above $+70^{\circ}$ 
or below $-70^{\circ}$) for follow-up observations.
\par
The sample, containing 28 GRBs, is presented in Table~\ref{sample.tab}. For 
each burst we have also listed the Sun-to-field distance 
($\theta_\mathrm{Sun}$), the Moon-to-field distance ($\theta_\mathrm{Moon}$) 
and the Moon illumination at the time of the burst. This is done to examine 
if these parameters affect the redshift determination significantly, e.g. a 
full Moon close to a burst location. This is of course difficult to quantify 
as the OA brightness also plays a part. For example, GRB\,050820A has a 
measured redshift even if it occurred during a full Moon and 
$\theta_\mathrm{Moon} = 34^{\circ}$. Therefore, we decided not to limit the 
sample further. This relatively ``clean'' sample has a very high redshift 
recovery rate of almost 60\% (16/28).
\par
Figure~\ref{z.fig} shows the redshift distribution of the 16 bursts with a 
reported redshift in our {\it Swift} sample\footnote{Note that the sample 
contains all {\it Swift} bursts with a reported redshift except GRB\,050126 
($z = 1.29$).}. Both the mean and the median is $z \approx 2.75$, more 
than twice as large as the corresponding numbers for pre-{\it Swift} bursts
(1.37 and 1.04, respectively, calculated from a sample of 42 bursts). A 
natural explanation for this increase is the lower trigger threshold of 
{\it Swift} compared to previous missions, giving rise to fainter ({\it Swift} 
events are on average 1.7\,mag fainter in the $R$-band at a similar epoch: 
Berger et al. \cite{edo05}) and higher redshift bursts. This is complemented 
by the accurate positions provided by {\it Swift} and the rapid response of 
a variety of telescopes aimed at redshift determinations. 
\par
This {\it Swift} sample is the most uniform to date and it is of interest
to compare its redshift distribution to models predicting the fraction of 
GRBs expected to occur at a given redshift. Natarajan et al. (\cite{priya}, 
hereafter N05) have modelled the expected redshift distribution for GRBs, 
utilising several models including those which follow the globally averaged 
star formation rate (model II), and those which scale according to the 
average metallicity of the Universe at a given redshift (model IV, see e.g. 
Fynbo et al. \cite{johan}; Fryer \& Heger \cite{fryer}). Gorosabel et al. 
(\cite{javier}, hereafter G04) have also carried out a similar exercise, 
where the GRB rate is assumed to be proportional to the star formation rate. 
These models are plotted in Fig.~\ref{z.fig}.
\par
It is remarkable how similar the observed {\it Swift} redshift distribution
is to the model predictions. Contrary to inferences that could be drawn from 
the pre-{\it Swift} redshift distribution, we can now reason that GRBs trace 
star form\-ation (see also Jakobsson et al. \cite{palli05}). However, with 
the available sample and the limited flux sensitivity of the \emph{Swift}/BAT 
for $z > 5$ bursts, it is currently not possible to determine if GRBs are 
unbiased tracers of star formation. For example, models II and IV from N05 
are nearly indistinguishable when comparing to the relatively small sample 
of 16 bursts. Note that although model II from N05 and the G04 model both 
presuppose that the GRB rate is proportional to the star formation rate, they 
use different assumptions regarding the poorly determined GRB luminosity 
function and the intrinsic spectral shape, explaining their difference in 
Fig.~\ref{z.fig}. 
\par
By including all the bursts in Table~\ref{sample.tab} we are able to 
constrain the number of bursts above a specific redshift. For example, 
7\%--40\% of the bursts are located at $z > 5$. The N05 and G04 predictions 
are 10\% and 2\%, respectively, suggesting that the GRB luminosity function 
parameters and/or the GRB spectral index assumed in N05 might be more 
appropriate. Bromm \& Loeb (\cite{bromm}) also predict that 10\% of the 
\emph{Swift} GRBs should originate at $z > 5$.
\section{Discussion \& Conclusions}
\label{dis.sec}
The mean redshift of our relatively unbiased {\it Swift} sample 
($z_\mathrm{mean} = 2.8$) is larger than the mean redshift of sub-mm 
galaxies ($z_\mathrm{mean} = 2.4$: Chapman et al. \cite{chapman}) and is 
similar to that of Type 2 AGNs ($z_\mathrm{mean} \sim 3$: Padovani 
et al. \cite{padovani}). With two $z > 5$ GRBs discovered within a space of 
a month, and primarily due to the spectroscopic redshift of $z = 6.29$ for
GRB\,050904 (Kawai et al. \cite{kawai}), we are finally accessing the GRB 
high-redshift regime. Are we starting to probe the era of Pop III stars? 
If the transition between the dark ages and the era of reionization 
occurred around $z \approx 6$--7 (see e.g. Miralda-Escud\'e \cite{miralda} 
for a review), the answer might be positive. However, Abel et al. 
(\cite{abel}) have calculated that at most one massive metal-free star 
forms per pre-galactic halo, and since the GRB progenitor needs to be a 
member of a close binary system in the collapsar scenario (e.g. Fryer et al. 
\cite{fryer99}; MacFadyen \& Woosley \cite{MacF}; Zhang et al. \cite{zhang}),
it seems unlikely that Pop III stars could end their lives as GRBs. 
Woosley \& Heger (\cite{woosley}) have proposed a non-binary possibility
in the collapsar scenario, introducing unusually rapidly rotating
massive stars. It is therefore conceivable that Pop III stars are GRB 
progenitors, although it is evident that the number of unknowns is currently 
too large to arrive at a concrete conclusion.
\begin{figure}
\centering
\resizebox{\hsize}{!}{\includegraphics[bb=85 374 540 696,clip]{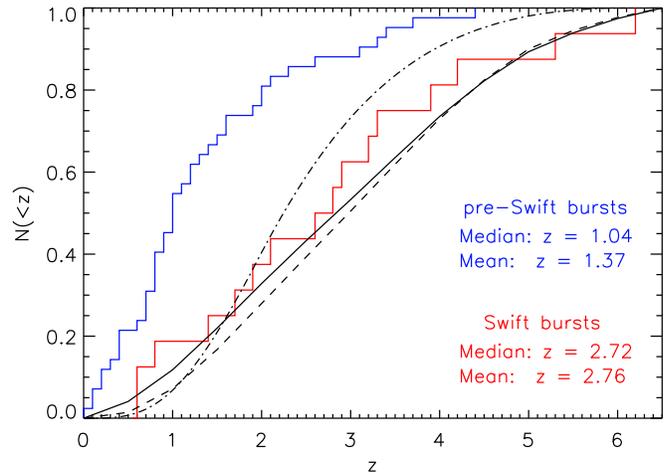}}
\caption{The cumulative fraction of GRBs as a function of redshift for 
42 pre-{\it Swift} bursts (upper stepwise curve) and 16 {\it Swift} bursts 
(lower stepwise curve). Overplotted are three simple models for the 
expectation of the redshift distribution of GRBs: model II from N05 in which 
the GRB rate is proportional to the star formation rate (solid curve), model 
IV from N05 in which the GRB rate increases with decreasing metallicity 
(dashed curve) and a model from G04 in which the GRB rate is proportional 
to the star formation rate (dash-dotted curve). All three models fold in 
the \emph{Swift}/BAT flux sensitivity.}
\label{z.fig}
\end{figure}
\par
The number of GRBs without an OA in our sample, sets a strict upper 
limit on the fraction of heavily dust obscured after\-glows as expected
in ultra\-luminous infrared host galaxies (ULIRGs). At most 20\% (6/28) of 
the GRB hosts in the sample can be of this nature. In fact, 5 of the 
non-detections only have a relatively shallow optical limit, typically 
$V \gtrsim 19$\,mag (albeit early post-burst), suggesting that the fraction 
is possibly lower. This is consistent with results from sub-mm observations 
of pre-\emph{Swift} burst (e.g. Tanvir et al. \cite{tanvir04}).
\par
The sample contains GRB\,050814, whose OA was particularly faint in the 
$R$-band; the observed optical to X-ray spectral slope is flatter 
($\beta_\mathrm{OX} = 0.36$) than expected for the fireball model. Hence, 
GRB\,050814 is classified as a dark burst as defined by Jakobsson et al. 
(\cite{palliDARK}). We have argued that this is most likely due to the 
high-redshift nature ($z = 5.3$) of this burst; the $R - I$ colour is 
extremely red which is impossible to explain by strong extinction given 
the observed $I - K$ colour. Indeed, a similar conclusion was proposed 
for GRB\,980329 (Fruchter \cite{andyZ}). 
\par
It is clear that GRBs have now opened up a window to the very high-redshift 
Universe. The emerging GRB redshift histo\-gram (Fig.~\ref{z.fig}) strongly
indicates that they can be used to trace the star formation in the 
Universe over a wide redshift range ($0 \lesssim z \lesssim 7$). Future 
instrumentation, such as the X-shooter (D'Odorico et al. \cite{xshooter}), 
will hopefully shed light on the end of the dark ages and the possible 
GRB/Pop III connection.

\begin{acknowledgements}
We thank Dale A. Frail and the anonymous referee for excellent comments. PJ, 
BLJ and KP acknowledge support from the Instrument Centre for Danish 
Astrophysics (IDA). RP acknowledges PPARC for support, while AL and NRT thank 
PPARC for support through postdoctoral and senior research fellowships, 
respectively. The research of JG is supported by the Spanish Ministry of 
Science and Education through programmes ESP2002-04124-C03-01 and 
AYA2004-01515. JMCC gratefully acknowledges partial support from IDA and the 
NBI's International Ph.D. School of Excellence. The Dark Cosmology Center is 
funded by the Danish National Research Foundation. The authors acknowledge 
benefits from collaboration within the EU FP5 Research Training Network 
``Gamma-Ray Bursts: An Enigma and a Tool". Based on observations made with 
the Nordic Optical Telescope (Isaac Newton Telescope), operated on the island 
of La Palma jointly by Denmark, Finland, Iceland, Norway, and Sweden (the 
Isaac Newton Group), in the Spanish Observatorio del Roque de los Muchachos 
of the Instituto de Astrof\'{\i}sica de Canarias.

\end{acknowledgements}


\begin{thebibliography}{}

\bibitem[2002]{abel} Abel, T., Bryan, G. L., \& Norman, M. L.
2002, Science, 295, 93

\bibitem[2005]{berger603} Berger, E., \& Becker, G. 2005, GCN Circ. 3520

\bibitem[2005]{berger318} Berger, E., \& Mulchaey, J. 2005, GCN Circ. 3122

\bibitem[2002]{edo02} Berger, E., Kulkarni, S. R., Bloom, J. S., et al.
2002, ApJ, 581, 981

\bibitem[2005a]{berger505} Berger, E., Cenko, S. B., Steidel, C., Reddy, N., 
\& Fox, D. B. 2005a, GCN Circ. 3368

\bibitem[2005b]{edo05} Berger, E., Kulkarni, S. R., Fox, D. B., et al.
2005b, ApJ, submitted (astro-ph/0505107)

\bibitem[2005]{bloom} Bloom, J. S. 2005, GCN Circ. 3990

\bibitem[2005a]{blustin801} Blustin, A., Band, D., Hunsberger, S., et al.
2005a, GCN Circ. 3733

\bibitem[2005b]{blustin} Blustin, A., Retter, A., Marshall, F.,
Chester, M., \& Gehrels, N. 2005b, GCN Circ. 3804


\bibitem[2005]{bromm} Bromm, V., \& Loeb, A. 2005, ApJ, submitted
(astro-ph/0509303)


\bibitem[2000]{calzetti} Calzetti, D., Armus, L., Bohlin, R. C., et al. 
2000, ApJ, 533, 682

\bibitem[2003]{cavanagh} Cavanagh, B., Hirst, P., Jenness, T., et al.
2003, ASPC, 295, 237

\bibitem[2005]{brad} Cenko, S. B. 2005, GCN Circ. 3807

\bibitem[2005a]{cenko502B} Cenko, S. B., Fox, D. B., Rich, J., et al.
2005a, GCN Circ. 3357

\bibitem[2005b]{cenko416A} Cenko, S. B., Kulkarni, S. R., Gal-Yam, A., \&
Berger, E. 2005b, GCN Circ. 3542

\bibitem[2003]{chapman} Chapman, S., Blain, A. W., Ivison, R. J., \&
Smail, I. R., 2003, Nature, 422, 695


\bibitem[2004]{lise} Christensen, L., Hjorth, J., \& Gorosabel, J.
2004, A\&A, 425, 913

\bibitem[2004]{xshooter} D'Odorico, S., Andersen, M. I., Conconi, P., et al. 
2004, SPIE, 5492, 220


\bibitem[2005]{foley} Foley, R. J., Chen, H.-W., Bloom, J., \& 
Prochaska, J. X. 2005, GCN Circ. 3483

\bibitem[1999]{andyZ} Fruchter, A. S. 1999, ApJ, 512, L1

\bibitem[1999]{andy} Fruchter, A. S., Pian, E., Thorsett, S. E., et al.
1999, ApJ, 516, 683

\bibitem[2005]{fryer} Fryer, C. L., \& Heger, A. 2005, ApJ, 623, 302

\bibitem[1999]{fryer99} Fryer, C. L., Woosley, S. E., \& Hartmann, D. H.
1999, ApJ, 526, 152

\bibitem[2005]{fugazza} Fugazza, D., Fiore, F., Patat, N., et al.
2005, GCN Circ. 3948

\bibitem[2001]{johan01} Fynbo, J. P. U., Gorosabel, J., Dall, T. H., et al.
2001, A\&A, 373, 796

\bibitem[2003]{johan} Fynbo, J. P. U., Jakobsson, P., M\o ller, P., et al.
2003, A\&A, 406, L63

\bibitem[2005a]{johan319} Fynbo, J. P. U., Hjorth, J., Jensen, B. L., et al.
2005a, GCN Circ. 3136

\bibitem[2005b]{johan802} Fynbo, J. P. U., Sollerman, J., Jensen, B. L., et al.
2005b, GCN Circ. 3749

\bibitem[2005c]{johan824} Fynbo, J. P. U., Jensen, B. L., Sollerman, J., et al.
2005c, GCN Circ. 3874

\bibitem[2004]{gehrels} Gehrels, N., Chincarini, G., Giommi, P., et al.
2004, ApJ, 611, 1005

\bibitem[2004]{ghirlanda} Ghirlanda, G., Ghisellini, G., Lazzati, D., 
\& Firmani, C. 2004, ApJ, 613, L13

\bibitem[2004]{javier} Gorosabel, J., Lund, N., Brandt, S., 
Westergaard, N. J., \& Castro Cer\'on, J. M. 2004, A\&A, 427, 87 (G04)

\bibitem[2005]{haislip} Haislip, J., Nysewander, M. C., Reichart, D., et al.
2005, Nature, submitted (astro-ph/0509660)

\bibitem[2003a]{jensNature} Hjorth, J., Sollerman, J., M\o ller, P., et al.
2003a, Nature, 423, 847

\bibitem[2003b]{hjorth} Hjorth, J., M\o ller, P., Gorosabel, J., et al. 
2003b, ApJ, 597, 699

\bibitem[2003]{stephen} Holland, S. T., Weidinger, M., Fynbo, J. P. U., et al.
2003, AJ, 125, 2291


\bibitem[2003]{palli03} Jakobsson, P., Hjorth, J., Fynbo, J. P. U., et al.
2003, A\&A, 408, 941

\bibitem[2004a]{palli04} Jakobsson, P., Hjorth, J., Fynbo, J. P. U., et al.
2004a, A\&A, 427, 785

\bibitem[2004b]{palliDARK} Jakobsson, P., Hjorth, J., Fynbo, J. P. U., et al.
2004b, ApJ, 617, L21

\bibitem[2005a]{palli05} Jakobsson, P., Bj\"ornsson, G., Fynbo, J. P. U., 
et al. 2005a, MNRAS, 362, 245

\bibitem[2005b]{palli-z} Jakobsson, P., Fynbo, J. P. U., Paraficz, D., et al.
2005b, GCN Circ. 4029


\bibitem[2001]{brian01} Jensen, B. L., Fynbo, J. U., Gorosabel, J., et al.
2001, A\&A, 370, 909

\bibitem[2005]{brian} Jensen, B. L., Fynbo, J. P. U., Hjorth, J., et al.
2005, GCN Circ. 3809

\bibitem[2005]{kawai} Kawai, N., Yamada, T., Kosugi, G., Hattori, T., \&
Aoki, K. 2005, GCN Circ. 3937

\bibitem[2005]{kelson} Kelson, D., \& Berger, E. 2005, GCN Circ. 3101


\bibitem[2000]{lamb} Lamb, D. Q., \& Reichart, D. E. 2000, ApJ, 536, 1

\bibitem[1992]{landolt} Landolt, A. U. 1992, AJ, 104, 340


\bibitem[1999]{MacF} MacFadyen, A. I., \& Woosley, S. E. 1999, ApJ, 524, 262

\bibitem[1995]{madau} Madau, P. 1995, ApJ, 441, 18

\bibitem[2003]{miralda} Miralda-Escud\'e, J. 2003, Science, 300, 1904

\bibitem[2005]{morris} Morris, D. C., Burrows, D. N., Kennea, J. A.,
Racusin, J. L., \& Gehrels, N. 2005, GCN Circ. 3805

\bibitem[2005]{mortsell} M\"ortsell, E., \& Sollerman, J. 
2005, JCAP, 6, 9

\bibitem[2005]{priya} Natarajan, P., Albanna, B., Hjorth, J., et  al.
2005, MNRAS, in press (astro-ph/0505496) (N05)

\bibitem[2004]{padovani} Padovani, P., Allen, M. G., Rosati, P., \& 
Walton, N. A. 2004, A\&A, 424, 545


\bibitem[1992]{pei} Pei, Y. C. 1992, ApJ, 395, 130

\bibitem[2003]{perna} Perna, R., Lazzati, D., \& Fiore, F. 
2003, ApJ, 585, 775


\bibitem[2005]{poole} Poole, T., Moretti, A., Holland, S. T., et al.
2005, GCN Circ. 3698

\bibitem[1995]{predehl} Predehl, P., \& Schmitt, J. H. M. M. 
1995, A\&A, 293, 889

\bibitem[2005]{price} Price, P. A., Cowie, L. L., Minezaki, T., et al.
2005, ApJ Letters, submitted (astro-ph/0509697)

\bibitem[2005]{jason} Prochaska, J. X., Bloom, J. S., Wright, J. T., et al.
2005, GCN Circ. 3833

\bibitem[2001]{reichart} Reichart, D. E. 2001, ApJ, 553, 235

\bibitem[2005]{retter} Retter, A., Barbier, L., Barthelmy, S., et al. 
2005, GCN Circ. 3799

\bibitem[2005]{rol} Rol, E., Schady, P., Hunsberger, S., et al.
2005, GCN Circ. 3186

\bibitem[1998]{schlegel} Schlegel, D. J., Finkbeiner, D. P., \& Davis, M.
1998, ApJ, 500, 525

\bibitem[1998]{spinrad} Spinrad, H., Stern, D., Bunker, A., et al.
1998, AJ, 116, 2617

\bibitem[2003]{stanek} Stanek, K. Z., Matheson, T., Garnavich, P. M., et al.
2003, ApJ, 591, L17

\bibitem[2005]{rhaana} Starling, R. L. C., Vreeswijk, P. M., Ellison, S. L.,
et al. 2005, A\&A, 442, L21

\bibitem[1992]{steidel92} Steidel, C. C., \& Hamilton, D. 1992, AJ, 104, 941

\bibitem[1993]{steidel93} Steidel, C. C., \& Hamilton, D. 1993, AJ, 105, 2017

\bibitem[2005]{tag} Tagliaferri, G., Antonelli, L. A., Chincarini, G., et al.
2005, A\&A Letters, in press (astro-ph/0509766)

\bibitem[2004]{tanvir04} Tanvir, N. R., Barnard, V. E., Blain, A. W., et al.
2004, MNRAS, 352, 1073

\bibitem[2005]{tanvir} Tanvir, N., Lowe, K., Gledhill, T., et al.
2005, GCN Circ. 3632

\bibitem[2005]{tueller} Tueller, J., Markwardt, C., Barbier, L., et al. 
2005, GCN Circ. 3803

\bibitem[2004]{paul} Vreeswijk, P. M., Ellison, S. L., Ledoux, C., et al.
2004, A\&A, 419, 927

\bibitem[2005a]{darach} Watson, D., Fynbo, J. P. U., Ledoux, C., et al.
2005a, ApJ, submitted

\bibitem[2005b]{darach904} Watson, D., Reeves, J. N., Hjorth, J., et al.
2005b, ApJ Letters, submitted (astro-ph/0509640)

\bibitem[1998]{weymann} Weymann, R. J., Stern, D., Bunker, A., et al.
1998, ApJ, 505, L95

\bibitem[2005]{woosley} Woosley S. E., \& Heger, A. 2005, ApJ,
submitted (astro-ph/0508175)



\bibitem[2004]{zhang} Zhang, W., Woosley, S. E., \& Heger, A.
2004, ApJ, 608, 365

\end{thebibliography}
\end{document}